# A Constellation of MicroSats to Search for NEOs

Michael Shao, Hanying Zhou, Slava G. Turyshev, Chengxing Zhai, Navtej Saini, and Russell Trahan

*Jet Propulsion Laboratory, California Institute of Technology,*
*4800 Oak Grove Drive, Pasadena, CA 91109-0899, USA*

## ABSTRACT

Large or even medium sized asteroids impacting the Earth can cause damage on a global scale. Existing and planned concepts for finding near-Earth objects (NEOs) with diameter of 140 m or larger would take ~15-20 years of observation to find ~90% of them. This includes both ground and space based projects. For smaller NEOs (~50-70 m in diameter), the time scale is many decades. The reason it takes so long to detect these objects is because most of the NEOs have highly elliptical orbits that bring them into the inner solar system once per orbit. If these objects cross the Earth's orbit when the Earth is on the other side of the Sun, they will not be detected by facilities on or around the Earth. A constellation of MicroSats in orbit around the Sun can dramatically reduce the time needed to find 90% of NEOs ~100-140 m in diameter.

## 1. INTRODUCTION

This paper is an extension of our previous effort to study the capabilities of microsatellites (MicroSats) for detection of the near-Earth objects (NEOs) which led to the conclusion that a constellation 6 MicroSats is capable of detecting ~90% of NEOs with the diameter of 140 m in ~ 3 years (Shao et al. 2017, Zhai et al., 2018).

Here we describe a more capable constellation of MicroSats, and evaluates not just the "detection" statistics but also the statistics for "cataloging" the newly found NEOs. When a new NEO is detected, a single detection provides no information about its orbit. If the initial detection is not followed up with subsequent detections in the next few days, the object will be lost, i.e., when it will be re-discovered again at a later time, one will not be able to link the two observations confirming that they relate to the same object. A cataloged NEO is one that is observed at least 3 times over the period of ~3 weeks. This set of three measurements result in a crude orbit such that a second cataloged observation several decades later can be linked to the first one.

We begin by briefly describing the technique of synthetic tracking (Shao et al. 2017). Synthetic tracking improves the signal-to-noise ratio (SNR) of an observation of a moving object by one or more orders of magnitude, enabling very small telescopes to have sensitivity equal to that of a much larger telescopes. Such a dramatic increase in sensitivity makes it possible to consider deploying a constellation of small telescopes that is not only much less expensive but also of a much higher performance than a single large ground- or space-based facility.

We then describe the simulation we performed using a constellation of 8 microsatellites relying on ~20 cm telescopes with large field of view (FOV) focal planes that can catalog 90% of 140 m NEOs in ~ 3 years of observation. While this simulation is interesting by itself, providing very valuable results, we also perform a range of simulations to understand why a constellation of small telescopes so vastly outperforms a single large telescope. These additional simulations provide a quantitative measure of the effect that we call saturation.

The simplest way to understand saturation is to think of a single large telescope in orbit that scans the sky with a sensitivity down to some faint limiting magnitude, say 23 mag, and can cover $4\pi$ steradians in 1 week. At 23 mag, a 140 m NEO (at opposition) can be detected at a distance up to 0.8 AU from the Earth. On average, these objects will move close enough to be detected and stay detectable for ~3 months until they are no longer brighter than 23 mag. What would be gained by placing a 2-nd such telescope into operation? The answer is - close to zero - because the 1-st telescope can scan the entire sky in 7 days, while the objects are detectable for an average of 120 days. Thus, the 1-st telescope is already $120/17 \approx 17$ times beyond saturation adding a 2-nd telescope at the same location will result in close to zero additional NEOs detected. At what distance would the 2-nd telescope have to be from the 1-st one to "avoid" saturation? The answer is -- very roughly 0.8 AU, which is the distance at which we can detect a 140 m NEO.

## 2. SYNTHETIC TRACKING

Detecting NEOs is different than detecting stationary objects. Traditionally NEO search telescopes take several ~45 sec CCD exposures over a time span of ~1 hr. Objects that appear to move linearly in time over that time span of ~1 hr are potential NEOs. The key is that the moving object be detected in each of the ~4 images taken. Because NEOs move, a long exposure results in a "streaked" image. The streaking results in spreading the photons over a larger number pixels making them harder to detect above the background sky noise. Synthetic tracking avoids this loss of SNR by combining multiple short exposures and then adding the image stack using a shift/add algorithm.

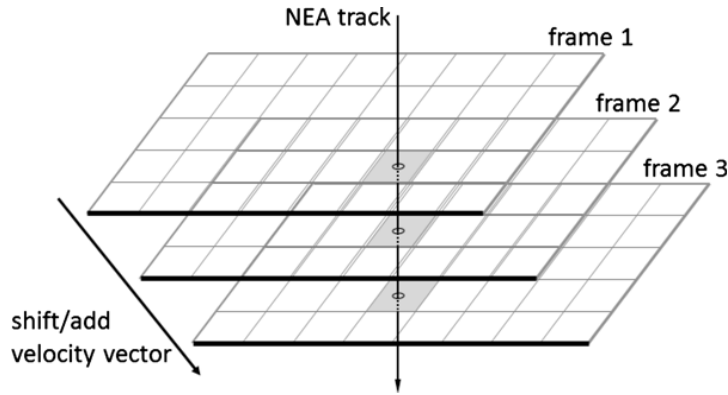

Figure 1. Shift-and-add concept illustrated: because of the motion of the NEA, photons are deposited on different pixels of a CCD, but in the synthetic image (with shifted/added frames) the asteroid smear is removed.

Synthetic tracking is possible because modern CMOS focal plane sensors can be readout at very low read noise (below photon noise from the sky background) at relatively fast frame rates compared to CCD sensors. Most large format CCD sensors that are used in a large mosaic focal plane needs over 10 sec to read out and also needs a mechanical shutter to prevent image smearing while reading the image out.

One apparent drawback to shift and add is that before we detect the object, we do not know in advance what it's velocity is and how far and in which direction we should shift subsequent images. Synthetic tracking does a brute force search, performing the shift/add in ~10,000 velocities. This is now possible with relatively low cost PCI-e board with GPUs that have peak processing speed of ~5 Tflop.

The use of synthetic tracking allows us to use very long integration times to make up for the smaller aperture telescope that we use. An example of a small telescope that would be uses for the search is a modified Schmidt telescope with a 28 cm aperture f/2.22 and a sensor with 3.8 μm pixels. A backside CMOS commercial sensor with 80% QE and < 2e– read-noise at 2 Hz will soon be available in a very large ~140 Mpix format. This would provide a FOV > 16 deg$^2$ and a limiting magnitude (in space) of ~22 mag assuming image quality better than 1.75 arcsec FWHM.

Table 1. Instrument input and derived parameters.

| Inputs parameters | Vaues | Units | Derived param. | Vaues | Units |
|---|---|---|---|---|---|
| NEO diameter | 140 | m | Apparent mag | 22.01 | mag |
| Distance | 0.615 | AU | Flux detected | 1.2 | e$^-$/sec |
| Transverse veloc. | 10 | km/sec | Noise/frame | 82.58 | variance e$^-$ |
| Phase angle | 0 | degrees | Signal/frame | 12.01 | e$^-$ |
| Telescope diam. | 279 | mm | | | |
| Total QE | 0.56 | | Sotal SNR | 7 | in 500 sec |
| Pixel size | 1.27 | " | | | |
| Read noise | 1.7 | e$^-$ | FOV | 17.79 | sqdeg |
| Frame time | 10 | sec | | | |
| Total Integ time | 500 | sec | | | |
| Sky background | 22 | mag/(")$^2$ | | | |

## 3. SIMULATION OF A CONSTELLATION OF 8 NICROSATS

Table 2 showed what is capable possible with the latest available hardware. Our simulations looked at a slightly less capable system. With a 10 sqdeg FOV and 22 mag limit in 600 seconds.

The simulation starts with a synthetic population of NEOs from (Granvik et al., 2016). We performed simulations of the entire NEO population as well as the "impactor" NEO population. By impactors we mean NEOs that have orbits which come within 0.01 AU of the Earth's orbit, or a MOID (minimum orbit intersection distance) of < 0.01 AU. We place the 8 telescopes evenly spaced in a ~1AU orbit around the Sun. Each satellite then systematically scanned the sky in an orange peel pattern. The simulation proceeded in steps of 600 seconds. At each step we calculated the position of the synthetic NEOs in the solar system as well as the 8 satellites. If a NEO was within the FOV of the telescope, we would calculate the apparent magnitude of the NEO as seen from the telescope. For observations, when the phase angle was not zero, we used the standard HG scattering function, with G = 0.15 that drops significantly more rapidly than Lambertian scattering with large phase angles. If the apparent magnitude was brighter than the limiting magnitude of the telescope/camera, it would be recorded as a detection. The simulation was then run continuously in 600 sec steps typically for ~6 years.

We varied a number of parameters in an attempt to optimize the number of NEOs detected. Because of the HG scattering function, NEOs at a fixed distance are much dimer at 90° phase angle than at 0° phase angle. Our search pattern had a variable sun-exclusion angle. We would skip over the parts of the sky where the telescope pointing was less than the Sun-exclusion angle. We also allowed the telescope's orbit around the Sun to be varied. We adjusted the sky background (zodi light) to reflect the fact that sky got brighter as the orbit radius got to be less than 1 AU.

Different constellations of MicroSats were investigated and we report only a small subset of the results. We collected statistics on three different types of detections. One is a simple or single detection. A second statistic was what we called a linked/cataloged detection. This is where the NEO is detected at least 3 times over a ~21 day interval. Or the detections allowed for an orbit solution that was as good as 3 observations evenly spaced over 21 days. The third statistic was equivalent to a reasonable orbit detection, where the same NEO was detected a minimum of 6 times over an orbital arc of at least 120° around the Sun.

Table 2. Simulation results.

| Pct of completion after 6 yrs | Case 1, 8 cubesats | | | Case 2, 4 paired cubesats | | | Case 3, 6 paired cubesats | | |
|---|---|---|---|---|---|---|---|---|---|
| | H =22 | H =23 | H =24 | H =22 | H =23 | H =24 | H =22 | H =23 | H =24 |
| One det | 98% | 87% | 61% | 85% | 69% | 47% | 95% | 82% | 60% |
| Linked det | 75% | 47% | 19% | 80% | 62% | 36% | 93% | 75% | 47% |
| Orbit det120 | | | | 36% | 17% | 6% | 48% | 25% | 11% |

Case 1 was 8 satellites uniformly spaced around the Sun. An H=22 mag NEO is ≈140 m in diameter. H=23 mag is 88 m and H = 24 is 56 m in diameter. We see in this case the completion rate for single detections was extremely high 98% after 8 years. However, the cataloged or linked completion rate was much lower only 75%. We did not collect statistics for the 3-rd category for Case 1. Case 2 also had 8 satellites but in 4 pairs. The pairs were space about 200,000 km apart so that we could get a parallax measurement of the distance to the NEO. This improved our link detection statistic from 75% to 80% for 140 m NEO impactors. Case 3 has a total of 12 satellites in 6 pairs. With 12 satellites we find that ~93% of 140 m NEOs would be cataloged in 6 years. 48% would have reasonably good orbits.

We should explain the difference between the three detection, linked detection and orbit detection in a bit more detail. If a NEO is only detected once, we know its brightness, position and velocity at one point in time. We do not know its orbit. If 10 years later that NEO is detected again, by a subsequent mission/facility we cannot link those two observations together. In some sense, an isolated detection of a NEO will result in that object subsequently being lost. The concept of a "linked" or cataloged observation is that there is now a sequence of measurements, from which we can derive some orbital information. Enough that a similar "cataloged" observation set a few decades later can be linked to the original set of observations.

Note that a cataloged observation is not sufficient to "predict" where in the Sky the NEO will be a few decades later. With ~3-4 observations within ~20 days, The derived orbital parameters are not very precise. In particular the semi-major axis of the orbit may be in error by, say, 1%. A 1% error in the semi-major axis means the orbit period has an error of ~1.5%. A few decades later, say 10 times the orbital periods, the uncertainty in the orbital phase or where the NEO is in its orbit is 15% of the circumference of the orbit.

Which would be an arc that extends over 1 AU. As seen from Earth, that 1 AU arc could extend ~60-120° across the sky.

Of the 6 orbital parameters, orbital phase degrades with time. But many of the other orbital parameters do not degrade with time. The pole of the orbit and the semi-major axis does not degrade with time. Two cataloged observation sets can be linked when the orbital parameters that don not degrade with time match. A cataloged observation set cannot be linked to a single future observation, but can be linked to a future "set" of observations that constitute a 2-nd cataloged observation.

When the NEO is observed over a short arc, orbital parameters are not well determined. In general the accuracy improves as the square of the arc length. As the length of the arc further increases to be larger than ~90° of motion around the Sun, the accuracy of the orbital parameters will improve linearly with the arclength. And, if the NEO is observed across multiple orbits, the improvement in accuracy grows as the square root of the number of observations. Our third statistic 120° orbital arc represents observations of a NEO by more than 1 satellite in the constellation in the region where orbital accuracy is now increasing linearly with arclength. Very roughly speaking this would allow a single observation of that NEO to be linked to the 120° orbit within a few years of the initial set of observations.

NEOCam and LSST have their simulation results published and as a comparison, after 6 years, for 140 m NEOs, NEOCam 82% single detections and 72% linked/cataloged (Mainzer et. al, 2016). For LSST, 60% cataloged/linked after 10 yrs of observation (Chesley et. al 2017).

## 4. SATURATION

For telescopes with a small FOV and rather low sensitivity, doubling the number of telescopes will double the discovery rate of NEOs and decrease by a factor of 2 the time needed to find 90%. If the sensitivity of the telescope were to increase by 1 mag, from say 20 mag to 21 mag, the distance a 140 m NEO could be detected at 0 phase angle would increase from 0.30 AU to 0.43 AU. This increase in range would increase the volume of search space by almost a factor of 3 and one might think this would increase the discovery rate by a factor of 3. However, as sensitivity and FOV increases, we see an effect we call saturation, that is the decrease in the time to detect 90% of the NEOs is much less than one would expect given the larger FOV or the higher sensitivity.

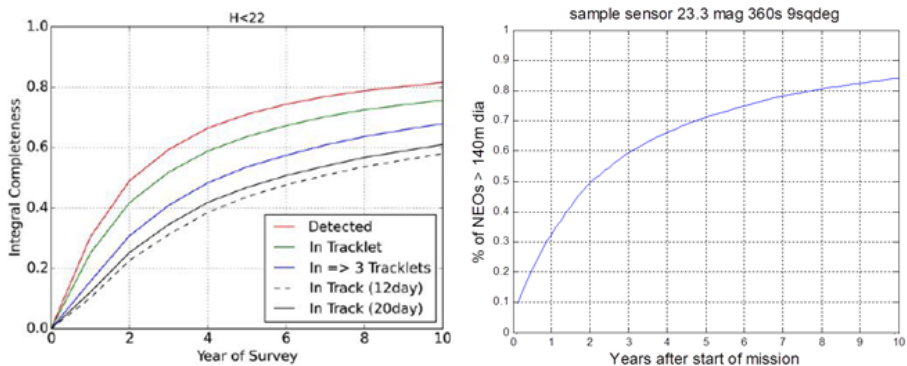

Figure 2. Left LSST detection of H<22mag NEOs versus time (Chesley 2017). Right a single telescope in space configured to have similar performance to LSST. This is subsequently scaled to examine saturation effects.

There are several ways to understand this. If 140 m NEOs were detectable out to 0.4 AU from Earth, and if on average their relative velocity to Earth is 10 km/s, then they will typically be detectable for ~0.3 AU/10km/s or ~50 days. If the telescope/camera can survey the night time sky (~20,000 sqdeg) in 20 days, Then building another telescope and doubling the sky coverage area would not double the rate of detection of new NEOs because the 2-nd telescope would just detect the NEOs found by the 1st. Most NEOs have highly eccentric orbits and come into the inner solar system (inside the orbit of Mars, for only a small portion of its orbit. If the Earth is on the other side of the Sun when that happens, they will not be detected, regardless of the sensitivity of the telescope. It would be useful to quantify the saturation effect, to find out how far into saturation the next generation of NEO search facilities would be.

Towards that goal, we constructed a model of a space based telescope operating in the visible band with synthetic tracking whose discovery rate is roughly that of LSST. The discovery rate of LSST was described in detail in a paper by Chesley et al. (2017). The %-age of NEOs larger than 140 m detected at least once is shown in the graph below. Alongside, is a space based telescope that has roughly the same performance. It is a system with a limiting magnitude of 23.3mag and can scan 9 sqdeg in 360 sec. Note that because this is in space it is able to observe 24 hrs a day and the moon and weather are not the limiting factors. The red line in the LSST plot is the single detection curve that should be compared to the surrogate on the right hand plot.

We then perturbed the "baseline" system first by doubling its sky coverage sqdeg/hr, equal to building two of these telescopes/cameras. Then we reduced that by 50%. The number of unique NEOs < H=22 mag versus time is plotted in the next two graphs.

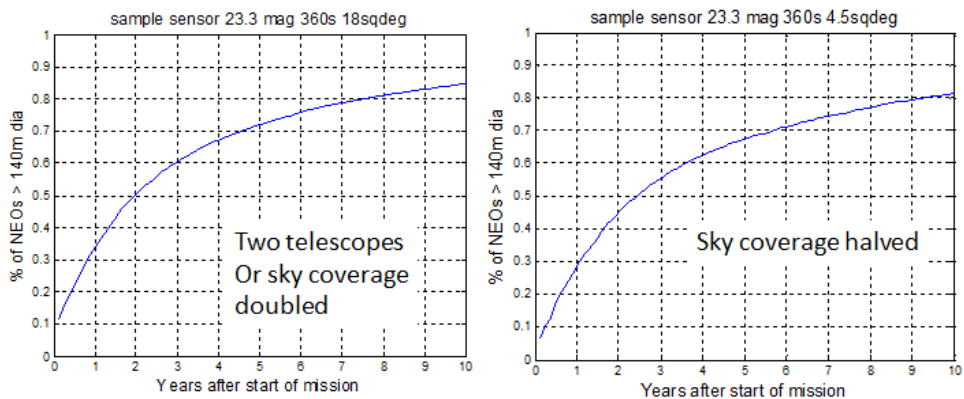

Figure 3. Left % of H<22mag NEOs detected versus time for two telescopes. Right % of H<22 mag for a telescope with ½ the sky coverage (in sqdeg/hr).

We see that adding a 2-nd telescope only increased the %-age of NEOs from ~83% to 84.8%. Similarly, halving the sky coverage only decreased it to 81.4%. Next, we considered increasing the sensitivity, by letting the size and cost of the telescope increase by a factor of two. Many cost models of telescope have the cost grow as ~diameter$^{2.5}$. Using this a telescope with twice the cost would have a 1.3X larger diameter and its sensitivity would be 0.28 mag better. We then also simulated putting a 2-nd such telescope in solar orbit on the other side of the Sun. These two are shown in the two plots in Figure 4. Last of all, the effects of saturation are shown in the Table 3.

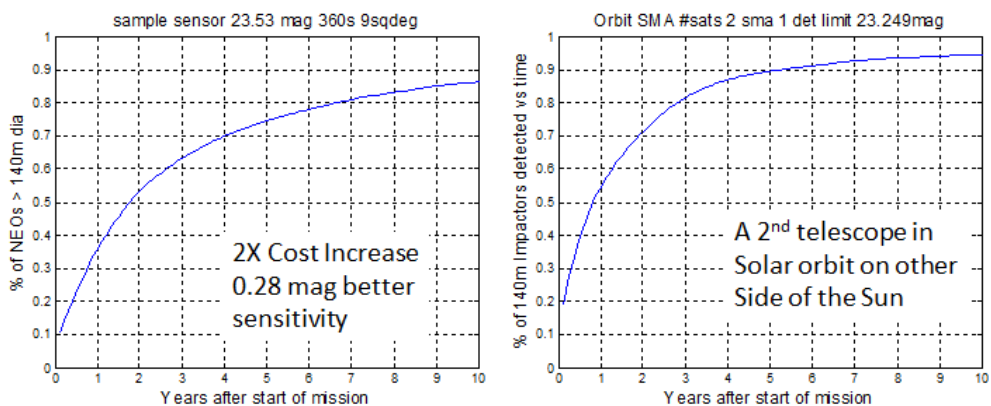

Figure 4. Left is the %-age of NEOs vs time for a telescope with ~1.3 times larger in diameter, 0.28 mag fainter limiting magnitude roughly doubles the cost (which scales as ~D$^{2.5}$).

Table 3. Summary of saturation effects

| Saturation Effect | % 10 yr | years @ 80% |
|---|---|---|
| Single large telescope | 83.8% | 7.8 |
| Two in Earth orbit | 84.8% | 7.5 |
| 1/2 in Earth orbit | 81.4% | 9 |
| 1.3X dia larger 0.28 mag | 86.0% | 6.5 |
| Two in solar orbit | 95.0% | 2.7 |

The column "% in 10 yrs" is the %-age of NEOs with diameter of > 140 m detected one or more times in 10 years. The column "years @ 80%" is the number of years it takes to detect 80% of the NEOs. From this simple example, it seems there might be a small advantage to building a larger telescope at twice the cost rather than building two telescopes. However that is based on the $D^{2.5}$ cost model. When building several telescopes the 2-nd is often less than building the 1-st. On the other hand irrespective of the cost, the gain in adding a 2-nd telescope is quite small if it is close to the 1-st. When the 2-nd telescope is placed on the other side of the Sun, the gain is quite a bit larger. Single detection of a NEO is not the appropriate metric, cataloged detection is the correct metric, but the purpose of this numerical experiment was to quantify the effect of saturation.

## 5. SUMMARY

Synthetic tracking enables a small telescope to achieve similar sensitivity for moving objects as a much large telescope. The reduction in cost make it possible to consider a constellation of MicroSats distributed around the Sun, that avoids the saturation effect that would makes it virtually impossible to catalog 90% of 140 m NEOs in less than 10 years.

This work in part was performed at the Jet Propulsion Laboratory, California Institute of Technology, under a contract with the National Aeronautics and Space Administration, and in part from a grant from the B612 Foundation. © 2018 California Institute of Technology. U.S. Government sponsorship acknowledged.